\documentclass{rmf-d}
\usepackage{nopageno,rmfbib,multicol,times,epsf,amsmath,amssymb,cite,upgreek,mathrsfs,mathtools}
\usepackage[latin1]{inputenc}
\usepackage[]{caption2}
\usepackage{graphicx}
\makeatletter
\newcommand{\vast}{\bBigg@{4}}
\newcommand{\Vast}{\bBigg@{5}}
\makeatother
\def\rmfcornisa{Research OR Education of Physics \hfill\rmf\ {\bf ??} (*?*) ???--???
\hfill MES? A\~NO?}

\def\rmfcintilla{{\it Rev.\ Mex.\ Fis.\/} {\bf ??} (*?*) (????) ???--???}
\clearpage \rmfcaptionstyle 
\begin{document}
\title{Ratios of Partial Wave Amplitudes in the Decays of $J=1$ and $J=2$ Mesons.
\vspace{-6pt}}
\author{Vanamali Shastry}
\address{Institute of Physics, Jan Kochanowski University, ul.  Uniwersytecka  7,  P-25-406  Kielce,  Poland}
\maketitle
\recibido{day month year}{day month year
\vspace{-12pt}}
\begin{abstract}
\vspace{1em} We study the two-body decay of mesons using the covariant helicity formalism. We find that to explain the ratio of partials wave amplitudes (PWAs) of the decay, the Lagrangian must include derivative interactions in addition to contact interactions. We estimate the ratios of the coupling constants for the vector decays of the axial-vector, pseudovector and pseudotensor meson, and the tensor decays of the pseudotensor mesons by fitting the ratios of the PWAs to the available data and predict the ratios to some new decays.\vspace{1em}
\end{abstract}
\keys{meson decay, partial wave amplitudes, nonlocal interactions \vspace{-4pt}}
\pacs{14.40.Cs, 12.40.-y, 13.30.Eg, 13.20.Jf \vspace{-4pt}}
\begin{multicols}{2}

\section{Introduction}\label{sec1}
Meson spectroscopy has been dealt with in great detail, using a multitude techniques in the past decades. In the light meson sector, the quark models have been successful not only in explaining the properties of the mesons, but also in predicting new states - both conventional and exotic, and their decay schemes\cite{Godfrey:1985xj,Barnes:1996ff}. Low energy hadronic effective field theories have been successful not only in studying the decays of the mesons but have also been used to study the electromagnetic properties of baryons. \cite{Parganlija:2012fy,Bernard:1995dp}.\par

Often, the coupling constants of the meson-meson interactions are estimated using the partial widths of the respective decays. To this end, most of the models employ Lagrangians that involve the lowest order interactions. When a meson of spin $J$ decays to a meson with the same spin and another spin-less meson, the lowest order interaction is the contact interaction. Such simple Lagrangian models are capable of explaining the width of the given decay. However, they do not capture the complexity and the structure of the interacting states.\par

In this work (based on Ref. \cite{Shastry:2021asu}) we show that even though the lowest order interactions reproduce the decay widths, are not sufficient to explain {\it all} the properties of a decay process. To this end, we analyse the partial wave structure of the decays of mesons using the covariant helicity formalism advocated by Chung \cite{Chung:1993da,Chung:1997jn,Filippini:1995yc}. We proceed to calculate the ratios of the various partial wave amplitudes (PWAs) and use them to estimate the ratios of the coupling constants.\par

This paper is organized into the following sections: in Sec. \ref{sec2}, we briefly discuss the partial wave analysis of two-body of decays; in Sec. \ref{sec3} we apply the covariant helicity formalism to the decays of $a_1(1260)$, $b_1(1235)$, and $\pi_2(1670)$. Finally we the summarize the paper in Sec. \ref{sec5}

\section{Partial wave analysis of two-body decay}\label{sec2}
The decay of a meson ($A$) into two (massive) states ($B$ and $C$) proceeds via the angular momentum channels according to the rules of the addition of the angular momenta. If $J_A$, $J_B$, and $J_C$ are the spins of the states involved, and $\ell$ is the relative angular momentum, then,
\begin{align}
    \vec{J_A} &= \vec{J_B} + \vec{J_C} + \vec{\ell}.
\end{align}
Since all the decays that we have studied involve one pseudoscalar meson as a product, $J_C$ can be taken to be $0$. Hence, the angular momentum can take the values $\ell=|J_A-J_B|,\ldots ,J_A+J_B$. However, not all these values of $\ell$ are allowed. The angular momentum must also satisfy the following relation,
\begin{align}
    (-1)^{P_A} &= (-1)^{\ell+1} (-1)^{P_B}
\end{align}
where, $P_A$ and $P_B$ are the parity quantum numbers of the states $A$ and $B$. The exponent $\ell+1$ is due to the angular momentum $\ell$ and the parity of the pseudoscalar meson. Thus, if the states $A$ and $B$ have the same parity, then only odd values of $\ell$ are allowed and vice versa. In the next subsection we discuss the conventional partial wave analysis using the spherical harmonics. In the subsequent discussions, we do not assume that $J_C=0$.\par

\subsection{Partial wave expansion}
To expand the decay amplitude in terms of the spherical harmonics, one should first decouple the angular momentum from the spin of the states involved. We note that the spin of the decaying meson is given by,
\begin{align}
    |J_A,M_A\rangle &= |S,m_s\rangle \otimes |\ell,m_\ell\rangle
\end{align}
where, $|S,m_s\rangle = |J_B,M_B\rangle \otimes |J_C,M_C\rangle$ is the spin of the two-body final state, and $|\ell,m_\ell\rangle$ represents the relative angular momentum carried by the decay products. The symbol $\otimes$ represents the outer product between the corresponding states. The addition of the spin and the relative angular momentum is carried out using the rules of the addition of the angular momenta. Accordingly, we get,
\begin{align}
    |J_A,M_A\rangle &= \sum_{m_s,m_\ell} |\ell,m_\ell,S,m_s\rangle\langle\ell,m_\ell,S,m_s|J_A,M_A\rangle
\end{align}
where, the summation over $m_s(m_\ell)$ runs from $-S(-\ell)$ to $S(\ell)$, and $\langle\ell,m_\ell,S,m_s|J_A,M_A\rangle$ represents the Clebsch-Gordan coefficient that relates the state $|J_A,M_A\rangle$ with the states $|\ell,m_\ell,S,m_s\rangle$. Thus, we can write the amplitude as \cite{Jeong:2018exh},
\begin{align}
    i\mathcal{M}(\theta,\phi;M_A)\! &= i \sum_\ell \sum_{m_\ell=-\ell}^\ell\!\! G_\ell^{SH}\langle \ell,m_\ell,S,m_s | J_A,M_A \rangle\nonumber\\
    &\qquad \times Y_{\ell m_\ell}(\theta,\phi)\label{conpwa}
\end{align}
where, $Y_{\ell m_\ell}(\theta,\phi)$ are the spherical harmonics, and the coefficients ($G_\ell^{SH}$) of the spherical harmonics are proportional to the amplitudes of the corresponding partial waves.\par
For the frame of reference where the decay products move along the $z$-axis, $\theta=0=\phi$. The Eq. \ref{conpwa} takes the form,
\begin{align}
    i\mathcal{M}(0,0;M_A)\! &= i \sum_\ell \sum_{m_\ell=-\ell}^\ell\!\! G_\ell^{SH}\langle \ell,m_\ell,S,m_s | J_A,M_A \rangle\nonumber\\
    &\qquad \times \sqrt{\frac{2\ell+1}{4\pi}}\delta_{m_\ell 0}
\end{align}
where we have used the relation $Y_{\ell m_\ell}(0,0)=\sqrt{\frac{2\ell+1}{4\pi}}\delta_{m_\ell 0}$. \par

\subsection{Covariant helicity formalism}
Alternatively, one can extract the PWAs from the decay amplitudes using the covariant helicity formalism \cite{Chung:1993da,Chung:1997jn,Filippini:1995yc}. To do so, we write the amplitude for the decay process as
\begin{align}
    i\mathcal{M}_{A\to BC}(\theta,\phi;M_A) &= D^{J_A*}_{M_AM_B}(\phi,\theta,0)~ F_{M_BM_C}^{J_A}\label{helamp}
\end{align}
where, $D^{J_A}_{M_AM_B}(\phi,\theta,0)$ is the Wigner $D-$matrix, and $F_{M_BM_C}^{J_A}$ are the helicity amplitudes. The helicity amplitude is related to the PWAs through the relation\footnote{The details of this derivation are given in the papers \cite{Chung:1993da,Chung:1997jn} and the report \cite{Chung:1971ri}.}
\begin{align}
     F_{M_B0}^{J_A} &= \sum_{\ell S} G_{\ell S}^{J_A}\sqrt{\frac{2\ell+1}{2J_A+1}}\nonumber\\
     &\times\langle\ell, 0 ,J_B,M_B-M_C|J_A,M_B-M_C\rangle\nonumber\\
     &\times\langle J_B, M_B, J_C,-M_C|S, M_B-M_C\rangle\label{helpwa}
\end{align}
where, $G_{\ell S}^{J_A}$ is the $\ell S-$coupling amplitude. It should be noted that the PWAs derived using the covariant helicity formalism follow the relation 
\begin{align}
\sum_\text{spin}|i\mathcal{M}_{A\to BC}|^2 &= \sum_\ell |G_{\ell S}^{J_A}|^2.
\end{align}
For the decays we have studied, the PWAs derived using the covariant helicity formalism are related to the ones derived using the conventional partial wave expansion through the relation,
\begin{align}
    G_\ell^{SH} &= \sqrt{\frac{4\pi}{2J_A+1}} G_{\ell S}^{J_A}.
\end{align}
In a theoretical study, since we are deriving the decay amplitude from the corresponding Lagrangian, helicity amplitudes would automatically be derived from the polarization tensors and hence, be Lorentz invariant. This is evident from the fact that the helicity amplitudes are dependent only on the ratio $E/m$ (or, equivalently, $k/m$), where $E$ and $m$ are the energy and mass of the decay product. Thus, it would be logical to use the covariant helicity formalism in a theoretical study of the partial wave structure of the decays. In the next section, we apply the covariant helicity formalism to the decays of the axial-vector, pseudovector, and pseudotensor mesons. In the subsequent discussions, weuse the notation $G_\ell$ to represent the PWAs $G_{\ell S}^{J_A}$. \par

\section{Application to meson decays}\label{sec3}
In this section, we apply the covariant helicity formalism to the strong decays of the $1^{++}$, $1^{+-}$, and $2^{-+}$ mesons. We restrict the discussion to the decays of the isovector mesons.
\subsection{J=1}
The Lagrangian describing the $b_1(1235)\to\omega\pi$ decay is given by,
\begin{align}
    \mathcal{L}&= ig_B^c \langle b_{1,\mu}\omega^\mu \pi\rangle + ig_B^d\langle\mathfrak{b}_{1,\mu\nu}\upomega^{\mu\nu}\pi\rangle\label{lagJ1}
\end{align}
where, $g_B^c$ and $g_B^d$ are the coupling constants, $\mathfrak{b}_{1,\mu\nu}=\partial_\mu b_{1,\nu}-\partial_\nu b_{1,\mu}$, $\upomega^{\mu\nu}=\partial^\mu \omega^{\nu}-\partial^\nu \omega^{\mu}$, and $\langle~\rangle$ represents trace over the isospin. The first term in the Lagrangian represents the (local) contact interactions. The second term represents the (nonlocal) derivative interactions. The amplitude for this decay can be written as,
\begin{align}
i\mathcal{M} &= i g_B^c~ \epsilon_\mu(0,M_A)\epsilon^{\mu\ast}(\vec{k},M_B)\nonumber\\
& + i 2g_B^d \left( k_0\cdot k_1 ~ \epsilon^\mu(\vec{0},M_A)\epsilon^\ast_\mu(\vec{k_1},M_B)\right.\nonumber\\
&\left. - k_0^\nu~ k_{1,\mu}~\epsilon^\mu(\vec{0},M_A)\epsilon^\ast_\nu(\vec{k_1},M_B)\right)\nonumber\\
    &= -i\Bigg\{ \begin{matrix*}[l]g_B^c + 2g_B^d~\!M_\omega~\!E_\omega& M_A=M_B=\pm 1\\ \gamma(g_B^c+2g_B^d~\! M_\omega~\! E_\omega\\ - 2g_B^d~\!M_\omega~\! \beta~\! k)& M_A=M_B=0 \end{matrix*}\label{bomega}
\end{align}
where, $\gamma=\sqrt{1-\beta^2}=\sqrt{1-k^2/M_\omega^2}$, $M_\omega$ and $E_\omega$ are respectively the mass and energy of the $\omega$, and $k$ is the 3-momentum of the $\omega$. The PWAs can be extracted using Eq. \ref{helamp} and Eq. \ref{helpwa}. We note here that the contact interaction alone can produce higher partial waves. From Eq. \ref{bomega}, we see that for $M_A=M_B=0$, the contact interaction gives a factor of $\gamma$. When the 3-momentum is smaller than the mass of the meson, $\gamma=1-k^2/(2M_\omega^2)+\ldots$. The presence of higher powers of 3-momentum implies the presence of the higher partial waves.\par

The experimental value of the ratio of the PWAs is $0.277\pm0.027$ \cite{Zyla:2020ssz}. Using this value, we obtain the ratio of the coupling constants to be $g_B^d/g_B^c = -0.687\text{ GeV}^{-2}$.
This value is in close agreement with the value obtained in Ref. \cite{Jeong:2018exh}. In the absence of the derivative interactions, the ratio of the PWAs is independent of the coupling constant, and takes the value $-0.043$. This value is much lower than the experimental value indicating that the derivative interactions are essential to describe the $b_1(1235)\to\omega\pi$ decay. Further, we observe that $\Big|g_B^d/g_B^c\Big| \approx 1/M_{b_1}^2$ indicating that the derivative interactions are as dominant as the contact interactions and the two interfere destructively during the decay.\par
An interesting deviation from this behavior is that of the $a_1(1260)\to\rho\pi$ decay. The Lagrangian describing this decay is similar to the Lagrangian given in Eq. \ref{lagJ1}. The masses of the mesons involved in both these decays and the 3-momentum carried by the decay products are nearly identical. Thus, one would expect the two decays to have similar $G_2/G_0$ ratios. However, the corresponding experimental value for the $a_1(1260)\to\rho\pi$ decay is $G_2/G_0=-0.062\pm 0.020$ \cite{Zyla:2020ssz}. This value can be obtained without using the derivative interactions. If we include the derivative interactions and analyse the decay, we find that the ratio of the coupling constants is $g_d^A/g_c^A = -0.082$ which is much smaller than $1/M_{a_1}^2$. This indicates that the derivative interactions are less significant in the $a_1(1260)\to\rho\pi$ decay.
\end{multicols}
\subsection{J=2}
In this subsection, we study the $\pi_2(1670)\to f_2\pi$ (tensor mode) and $\pi_2(1670)\to\rho\pi$ (vector mode) decays. To study the tensor mode we use the Lagrangian,
\begin{align}
    \mathcal{L}_{f_2\pi} &= \cos\theta_T\left(g_{PT}^c\langle\pi_{2,\mu\nu}f_2^{\mu\nu}\pi\rangle + g_{PT}^d\langle\uppi_{2,\alpha\mu\nu}\mathfrak{f}_2^{\alpha\mu\nu}\pi\rangle\right)\label{lagpi2}
\end{align}
where, $\uppi_{2,\alpha\mu\nu}=\partial_\alpha \pi_{2,\mu\nu}-\partial_\mu \pi_{2,\alpha\nu}$, $\mathfrak{f}_2^{\alpha\mu\nu}=\partial^\alpha f_2^{\mu\nu}-\partial^\mu f_2^{\alpha\nu}$, and $\theta_T (=5.7^\circ)$ is the angle of mixing between the $2^{++}$ iso-singlets \cite{Dudek:2011tt}. The decay amplitudes are given by,
\begin{align}
i\mathcal{M}&=i\cos\beta_t\left(g_{PT}^c\epsilon_{\mu\nu}(\vec{0},M_A)\epsilon^{\mu\nu\ast}(\vec{k},M_B)+\! 2g_{PT}^d\left(k_0\cdot k_1\epsilon_{\mu\nu}(\vec{0},M_A)\epsilon^{\mu\nu\ast}(\vec{k_1},M_B)\!-\!k_{0,\alpha} k_{1}^\nu\epsilon_{\mu\nu}(\vec{0},M_A)\epsilon^{\alpha\mu\ast}(\vec{k_1},M_B)\right)\right)\label{ampJ2Ten}\\
&=\!i\cos\beta_t\Vast\{\begin{matrix*}[l]g_{PT}^c\dfrac{(M_{f_2}^2+2E_{f_2}^2)}{3M_{f_2}^2} +2 g_{PT}^d \dfrac{M_{\pi_2}}{M_{f_2}^2} E_{f_2} & M_A=M_B=0\\g_{PT}^c \dfrac{E_{f_2}}{M_{f_2}} + g_{PT}^d \dfrac{M_{\pi_2} }{M_{f_2}}(k^2+2M_{f_2}^2) & M_A=M_B=\pm1\\g_{PT}^c+2g_{PT}^dM_{\pi_2} E_{f_2}& M_A=M_B=\pm 2. \end{matrix*}\label{dertenamp}
\end{align}

The PWAs for this decay can be extracted similar to the case of the decay of the $J=1$ mesons. The PDG lists $G_2/G_0=-0.18\pm0.06$ for the tensor mode \cite{Zyla:2020ssz}. Using this value, we find that the ratio of the coupling constants is, $g_{PT}^d/g_{PT}^c = -0.209\text{ GeV}^{-2}$.
In addition to the $S-$wave and the $D-$wave, the tensor decay mode also involves the $\ell=4 (G)$ wave. However, the amplitude of the $G-$wave is very small compared to the $S-$wave. We obtain the ratio of the ratio of the corresponding PWAs as,
\begin{align}
    \frac{G_4}{G_0} &= 0.0042 \pm 0.0014.
\end{align}
Also, the $G-$waves have the same phase as the $S-$waves. Using the values of the coupling constants obtained, we can derive the ratios of the PWAs for the $\pi_2(1670)\to f_2'\pi$ decay. We get the values,
\begin{align}
   \frac{G_2}{G_0} = 0.0093\pm 0.0031, & \quad\frac{G_4}{G_0} = -(7.49\pm 2.7)\times 10^{-6}.
\end{align}
The very small values for the ratios is due to the fact that the 3-momentum carried by the decay products is very small.\par
We now move on to the vector mode of the decay of $\pi_2(1670)$. We use the Lagrangian given by,
\begin{align}
\mathcal{L}_{\rho\pi} &=ig_{PT}^v\langle\pi_{2,\mu\nu}\rho^\mu\partial^\nu\pi\rangle + ig_{PT}^t\langle\uppi_{2,\alpha\mu\nu}\uprho^{\alpha\mu}\partial^\nu\pi\rangle\label{lagpi2vec}
\end{align}
to study this decay. Here, $g_{PT}^v$ and $g_{PT}^t$ are the coupling constants for the vector and tensor iteractions respectively and $\uprho^{\mu\nu}=\partial^\mu\rho^\nu-\partial^\nu\rho^\nu$. The decay amplitude is given by,
\begin{align}
    i\mathcal{M} &= -g_{PT}^v~\epsilon_{\mu\nu}(\vec{0},M_A)\epsilon^{\mu\ast}(\vec{k_1},M_B)k_2^\nu-g_{PT}^t~\Big[2k_0\cdot k_1\epsilon_{\mu\nu}(\vec{0},M_A)\epsilon^{\mu\ast}(\vec{k_1},M_B)k_2^\nu - 2 k_{0,\mu}k_1^\alpha\epsilon_{\alpha\nu}(\vec{0},M_A)\epsilon^{\mu\ast}(\vec{k_1},M_B)k_2^\nu\Big]\label{ampJ2Vec} \\
    &= \frac{k}{\sqrt{2}} \Bigg\{\begin{matrix*}[l] (g_{PT}^v +2 g_{PT}^t M_{\pi_2} E_\rho) &\! M_A=M_B=\pm 1\\
    \dfrac{2}{\sqrt{3}}\left(\dfrac{E_\rho}{M_\rho} g_{PT}^v+ 2 g_{PT}^t M_{\pi_2} M_\rho\right)&\! M_A=M_B=0.\end{matrix*}
\end{align}
For this decay, the PDG lists the ratio of the PWAs as $G_3/G_1=-0.72\pm0.16$ \cite{Zyla:2020ssz}. This is a rather high value indicating that the $\ell=3$ ($F$) wave is nearly as strong as the $\ell=1$ ($P$) wave. The ratio of the coupling constants is $g_{PT}^t/g_{PT}^v = -0.255\text{ GeV}^{-2}$.
We see that the ratios of the coupling constants for tensor mode as well as the vector mode of the decay of the $\pi_2(1670)$ are of the order of $1/M_{\pi_2}^2$. Using this ratio of the coupling constants, we estimate the ratio of the PWAs for the $\pi_2(1670)\to K^*K$ decay as,
\begin{align}
    \frac{G_3}{G_1} &= -0.447 \pm 0.099.
\end{align}
The decays of the $\pi_2(1670)$ also exhibit properties similar to those for the decay of $b_1(1235)$, {\it viz.} the lower order interactions are insufficient to reproduce the experimental values of the ratios of PWAs, and the higher order interactions play equally important role in the decay process.\par
\begin{multicols}{2}
\section{Summary and Outlook}\label{sec5}
Summarizing, we have studied the partial wave structure of some of the decays of the axial-vector, pseudovector, and pseudotensor mesons. We find that higher order (nonlocal) interactions are necessary in order to explain the experimental data on the partial wave structure of the two-body decays of these mesons. The lower order interactions and the higher order interactions interfere destructively as indicated by the relative minus sign between the coupling constants. The higher order interactions contribute as much as the lower order interactions to the decays. Also, the contributions of the higher partial waves decrease as the 3-momentum carried by the decay product decreases.\par

The formalism described in this paper can be extended to the decays of the kaons and the isoscalars of the nonets. The mixing between the isoscalars due to the axial anomaly is an interesting topic in itself \cite{LHCb:2013ged,BESIII:2018ede,WA102:1997gkz,Koenigstein:2016tjw,Klempt:2007cp}. The behavior of the meson nonets under the $SU(3)_L\times SU(3)_R$ transformations seems to be related to the extent of the mixing between the isoscalars necessitated by the axial anomaly \cite{Giacosa:2017pos}. The most common theoretical approach to derive the mixing angle is to use the Gell-Mann-Okubo (GMO) mass relations. However, this leads to situations where the isoscalar mixing angles become dependent on the corresponding kaonic mixing angles, as in the case of the $1^{++}$ and $1^{+-}$ isoscalars \cite{Cheng:2011pb}. In the absence of precise information on the kaonic mixing angle \cite{Tayduganov:2011ui}, the isoscalar mixing angles extracted from the GMO relations become unreliable. Further, the GMO relations ignore the loop corrections, which can give large contributions if the meson is a broad state ({\it e.g.,} $a_1(1260)$, $h_1(1170)$, $K_1(1400)$, etc). An alternative is to study the decays of these isoscalars. Since the non-strange components of the isoscalars decay predominantly to isovector states and the strange component to the kaonic states, the ratio of the partial widths of these two decay channels will be proportional to functions of the mixing angle. The angle of mixing between the isoscalars of the $1^{++}$, $1^{+-}$ and $2^{-+}$ nonets have been studied by incorporating the higher order interactions in Ref. \cite{Shastry:2021asu}. This can be extended further to the study of the isoscalars of the other nonets.\par

The decays of other nonets can also be studied using this formalism. For example, the PDG lists the $D/S-$ratio for the decay of the exotic $\pi_1(1600)$ to $b_1(1235)\pi$ \cite{Baker:2003jh}. The rather high value of this ratio indicates that derivative interactions are needed to fully explain the properties of this state. The ratio of PWAs for the decays of the higher $J$ mesons can also help in constraining the parameters of their interactions and decays \cite{Wang:2016enc,Jafarzade:2021vhh}. Similar data exist in the baryon sector also. An in depth study of the partial wave structure of the decays of hadrons in general can reveal vital information about their nature and interactions.

\section*{Acknowledgements}
The work reported in Ref. \cite{Shastry:2021asu} was done in collaboration with Enrico Trotti and Francesco Giacosa. This work was supported by the National Science Centre (NCN) Poland via the OPUS project 2019/33/B/ST2/00613.

\end{multicols}
\medline
\begin{multicols}{2}

\end{multicols}
\end{document}